\documentclass[aps,pra,floatfix,showpacs,tightenlines,twocolumn,superscriptaddress]{revtex4}
\usepackage{times}
\usepackage{graphicx}
\usepackage{epsfig}
\begin{document}

\title{Weak non-linearities and cluster states}

\author{Sebastien G.R. Louis}\email{seblouis@nii.ac.jp}
\affiliation{National Institute of Informatics, 2-1-2
Hitotsubashi, Chiyoda-ku, Tokyo 101-8430, Japan}
\affiliation{Department of Informatics, School of Multidisciplinary Sciences,
The Graduate University for Advanced Studies,
2-1-2 Hitotsubashi, Chiyoda-ku, Tokyo 101-8430 Japan}

\author{Kae Nemoto}
\affiliation{National Institute of Informatics, 2-1-2
Hitotsubashi, Chiyoda-ku, Tokyo 101-8430, Japan}

\author{W. J. Munro}
\affiliation{Hewlett-Packard Laboratories, Filton Road, Stoke
Gifford, Bristol BS34 8QZ, United Kingdom}
\affiliation{National Institute of Informatics, 2-1-2
Hitotsubashi, Chiyoda-ku, Tokyo 101-8430, Japan}

\author{T. P. Spiller}
\affiliation{Hewlett-Packard Laboratories, Filton Road, Stoke
Gifford, Bristol BS34 8QZ, United Kingdom}

\date{\today}

\begin{abstract}
We propose a scalable approach to building cluster states of matter
qubits using coherent states of light. Recent work on the subject
relies on the use of single photonic qubits in the measurement
process. These schemes have a low initial success probability and
low detector efficiencies cause a serious blowup in resources. In
contrast, our approach uses continuous variables and highly
efficient measurements. We present a two-qubit scheme, with a simple
homodyne measurement system yielding an entangling operation with
success probability 1/2. Then we extend this to a three-qubit
interaction, increasing this probability to 3/4. We discuss the
important issues of the overhead cost and the time scaling, showing
how these can be vastly improved with access to this new probability
range.
\end{abstract}


\maketitle

\section{Introduction}

The intriguing idea of one-way or cluster state quantum computing
was initially developed by Briegel and Raussendorf \cite{rau01}.
They showed that a two-dimensional array of qubits, entangled in a
particular way (through Conditional Phase gates), combined with
single qubit operations, feed forward and measurements are
sufficient for universal quantum computation. All the required
interactions are already contained inside the system, and the
computation proceeds through a series of local measurements (with
classical feed forward), efficiently simulating quantum circuits. In
effect, the logical gates are prepared off-line and imprinted onto
the qubits as they are transmitted through the cluster.

This approach was quickly applied \cite{nie04,bro05,gilbert} to
linear optics quantum computing \cite{KLM01,Kok06} (and references therein), both having been
experimentally demonstrated \cite{zeil,kies}. It pushes the problem with the probabilistic nature of
2-qubit gates into the off-line preparation of the cluster
\cite{nie04,bro05}. In this context it was shown that simple parity
gates are sufficient for building the required states. These schemes
are then bounded by the single beam-splitter success probability of
1/2 and in fact this initial probability is far reduced when the
single photon detection inefficiencies are taken into account.
Supplementing the linear optical approaches with weak
nonlinearities\cite{mun05a,mun05b,nem05,yamaguchi06,spi06} allows for the
construction of parity gates with significantly higher success
probabilities (near unity in some cases). A core issue however with
photonic qubits is their `flying' nature and the storage
requirements this mandates.

A natural way around this issue is to move to solid state or
condensed matter qubits and use single photons for communication
between them. Many proposals make use of single photons to
effectively mediate interactions between matter qubits
\cite{bos99,cab99,fen03,dua03,bro03,sim03}. Having interacted with
them, the photons then interact with each other in a linear optical
setup before being measured, thus projecting the matter qubits into
the required state without them having interacted directly. It has
been shown that entanglement and logical operations can be generated
in this way. The next step was to use these probabilistic entangling
schemes to prepare cluster states of matter qubits
\cite{bar05a,lim06} using techniques like double-heralding or
repeat-until-success. However the schemes are generally limited by
the detection of the single photons (more than one in some cases).
This can severely limit the probability of realizing the entangling
operation and hence the creation of the cluster state. An
alternative is available and this is what we will describe in this
letter. Instead of using single photons, we can use coherent states
of light (similar to the weak nonlinearity approach). Homodyne
measurements on coherent light fields can be made much more
efficient than single photon detection and so we can achieve
entangling operations with a probability greater than 1/2. In this
paper we will show how this and other factors make continuous
variables a very powerful tool for growing matter qubit based
cluster states.

\section{Gates}

There are quite a number of well studied systems where one has a
natural interaction between the matter qubit and the electromagnetic
field. These include atoms (real and artifical) in CQED (both at the
optical and telecom wavelengths) \cite{raimond01}, NV-centers in
diamond \cite{jelezko04}, quantum dots with a single excess electron
\cite{pazy03}, trapped ions \cite{cir95} and SQUIDs
\cite{Shnirman97} to name only a few. All these systems are likely
to be suitable candidates for what we describe below but to
illustrate the details a little more clearly let us consider a
lambda based CQED system. One could use cesium atoms or an
NV-diamond center embedded in the cavity. Both of these systems
operate in the optical frequency range and so are well matched to
efficient homodyne measurements. The interaction between the
coherent field mode and our matter qubit can generally be described
by the Jaynes-Cummings interaction $\hbar g (\sigma^{-}a^{\dagger} +
\sigma^{+}a)$ and in the dispersive limit (large detunings) one
obtains an effective interaction Hamiltonian of the form
\cite{dispersive2,bla04}:
\begin{equation}
 H_{int}=\hbar\chi\sigma_z a^{\dag} a.
\label{1}
\end{equation}
where $a$ ($a^{\dagger}$) refers to the annihilation (creation) operator
of an electromagnetic field mode in a cavity and the matter qubit is described
using the conventional Pauli operators, with the computational basis being
given by the eigenstates of $\sigma_z\equiv |0\rangle \langle 0|- |1\rangle \langle 1|$,
with $|0\rangle \equiv |\uparrow_z\rangle$ and $|1\rangle \equiv |\downarrow_z\rangle$.
The atom-light coupling strength is determined via the parameter $\chi=g^2/\Delta$,
where $2g$ is the vacuum Rabi splitting for the dipole transition and $\Delta$ is
the detuning between the dipole transition and the cavity field. The interaction
$H_{int}$ applied for a time $t$ generates a conditional phase-rotation $\pm\theta$
(with $\theta=\chi t$) on the field mode dependent upon the state of the
matter qubit. We call this a {\it conditional rotation} and it is very similar to the
cross-Kerr interaction between photons. This time dependent interaction
implicitly requires a pulsed probe.

Now the interaction in (\ref{1}) forms the basis for an entangling
operation. A two-qubit gate has been proposed \cite{spi06} based on
controlled bus rotations and a subsequent measurement. The probe
field coherent state $|\alpha\rangle$  interacts with both qubits,
so an initial state of the system
$|\Psi_i\rangle=\frac{1}{2}(|00\rangle+|01\rangle+|10\rangle+|11\rangle)|\alpha\rangle$
evolves to
\begin{eqnarray}
|\Psi_f\rangle&=&\frac{1}{2} (|00\rangle|\alpha e^{2i\theta}\rangle+|11\rangle|\alpha
e^{-2i\theta}\rangle) \nonumber \\
&+& \frac{1}{2} (|01\rangle+|10\rangle)|\alpha\rangle.
\label{3}
\end{eqnarray}
Here we quickly observe that the probe field has evolved into three
potentially distinct states and appropriate measurements can project
our two qubits into a number of interesting states. At this stage we
can choose from different types of measurements on the probe beam.
The first and simplest option we have is to perform a homodyne
measurement of some field quadrature $X(\phi)=(a^{\dagger} e^{i
\phi} + a e^{-i \phi})$ which for a sufficiently strong local
oscillator (compared to the signal strength) implements a projective
measurement $|x(\phi)\rangle \langle x(\phi)|$ on the probe state
\cite{tyc04}. The key advantages with homodyne measurement, at least
in the optical regime are that it is highly efficient (99\% plus
\cite{polzik92}) and is a standard tool of continuous variable
experimentalists. The simplest homodyne measurement to perform is
the momentum ($P = X(\pi/2)$) quadrature. In this case the
measurement probability distribution has three peaks with the
overlap error between them given by
$P_{err}=\frac{1}{2}\text{erfc}(\alpha\sin \theta /\sqrt{2})$. As
long as $\alpha \theta \sim \pi$ this overlap error is small
($<10^{-3}$) and the peaks are well separated. If our $P$ quadrature
measurement projects us onto the central peak $|\alpha\rangle$, our
two matter qubits are conditioned into the entangled state
$(|01\rangle+|10\rangle)/\sqrt{2}$. This occurs with a probability
of 1/2. Detecting either of the other two side peaks will project
the qubits to the known product states $|00\rangle$ or $|11\rangle$.
The probability of entangling the two qubits is interesting in that
we have already reached the limits of conventional linear optical
implementations. When realistic detector efficiencies ($\eta\sim
70\%$) are taken into account, their optimal success probability of
$1/2$ decreases dramatically (proportional to $\eta$ or $\eta^2$
depending on the implementation) and so the probability of the
operation succeeding is now significantly less than 1/2. In contrast
homodyne measurements are highly efficient and so our success
probability will remain very close to 1/2. This limit may be
fundamental to the linear optical schemes but in our case we can
exceed it by changing the nature of our measurement. In principle we
could achieve a near deterministic gate if we  measured the the
position quadrature ($X=X(0)$), however the requirements to ensure
the distinguishability of the probe beam states are much more
severe. We could also in principle use a photon number measurement
after displacing the probe beam \cite{spi06}, but we would fall back
into the issues affecting the linear optical schemes. By restricting ourselves to $P =
X(\pi/2)$ quadrature measurements and single interactions 
between the qubits and the probe, we are opting for the most 
robust weak-nonlinear approach so far proposed.

Within the same framework of conditional rotations and $P$ 
measurements, one can envisage
three qubits interacting with a single probe beam. GHZ states are for instance
one particularly useful state \cite{bro05}. One way of projecting the
qubits onto GHZ-type states is to vary the strength of the interactions
between the qubits and the probe beam \cite{yamaguchi06}. Let us represent a rotation
of the coherent probe beam by 
$R(\theta \sigma_z) = \exp(i \theta a^{\dagger}a \sigma_z)$.
The sequence $R(\theta\sigma_{z_{1}})R(\theta\sigma_{z_{2}})
R(-2\theta\sigma_{z_{3}})|\alpha\rangle$ which we depict in Fig (1) will give the
optimal paths and end points in phase space.
\begin{figure}[!htb]\label{fig1}
\begin{center}\includegraphics[scale=0.5]{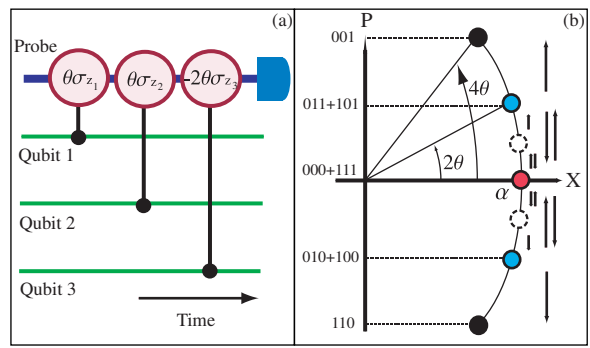}\end{center}
\caption{Schematic diagram (a) of a three qubit entangling operation. In (b)
the possible probe trajectories caused by the three conditional
rotations. There are five different end-states. Upon measurement, three
of these will project the qubits to entangled states of interest.}
\end{figure}
The peak centered on the origin will then correspond to the GHZ
state $(|000\rangle+|111\rangle)\sqrt{2}$ (after being detected).
This will happen with a probability of 1/4 (all qubits started in perfect superpositions). Next the two peaks
having been rotated through $\pm2\theta$ will correspond to the
qubit states $(|01\rangle_{1,2}+|10\rangle_{1,2})|1\rangle_{3}/\sqrt{2}$
and $(|01\rangle_{1,2}+|10\rangle_{1,2})|0\rangle_{3}/\sqrt{2}$
respectively. Now in both of these possible outcomes we obtain
the same Bell state on qubits 1 and 2, disentangled from qubit 3.
So overall we obtain a GHZ state with probability of 1/4 and a
Bell state with probability of 1/2 (on two qubits of our choice),
heralded by the probe beam $P$ quadrature measurement outcome.
The other two outcomes project the qubits to known product states. Consequently, if all we
want to do is entangle a pair of qubits, we can now do this with a
probability of 3/4.

This method can be extended to larger numbers of qubits, but the 3-qubit case minimizes the
ratio of operation time over success probability. We shall use this result in the remainder of the paper, observing how current work on the generation of cluster states is simply inadequate for probabilities exceeding 1/2. Until now strategies have been said to be scalable if the resources don't scale exponentially with the size of the cluster (in general they will scale sub-exponentially). This is a purely theoretical notion which bares little relation to the practical scalability we obtain in our approach.

We stress that although the 3-qubit operation is a probabilistic
entangling operation with different outcomes, these outcomes are 
heralded by the measurement of the bus and so the operation is a
very useful entangling primitive for the construction of cluster states.
For example, applying it to join two sections of cluster with a third ancillary
qubit works with probability 3/4, giving (heralded) outcomes of 
joined clusters with a new dangling bond (probability 1/2), or 
joined clusters and two new dangling bonds (probability 1/4). 
Applying the operation to join three sections of cluster
gives (heralded) outcomes of two sections joined and a new dangling bond
(probability 1/2), or all three sections joined and two new dangling
bonds (probability 1/4). All these outcomes contribute to cluster state
construction.

\section{Scaling}

Now let's turn to the issue of building up linear cluster states
(chains). In order to efficiently grow a chain with probabilistic
gates, one needs to first inefficiently build small chains exceeding
a critical length $L_c=1+2(1-p)/p$ and then try joining them to the
main one. This critical length varies between different entangling
operations. If an actual conditional phase gate can be immediately
implemented, then $L_c=2(1-p)/p$ for example. Or if this logical
gate requires the qubits from the cluster to interact directly
(non-distributive approach) then $L_c=4(1-p)/p$ \cite{dua05}.
Starting from this, and adopting a `divide and conquer' approach to
building these minimal chains, scaling relations are obtained for
the average number of entangling operations required and the average
time taken, to build a chain of length $L$. Using our 2-qubit gate
($L_c=3$) and these scaling relations we obtain $N[L]=12L-38$. This
is already the limiting scenario for simple single photon
applications. In the repeat-until-success method \cite{lim06}, for a
failure probability of 0.6 (and equal success and insurance
probabilities, on all results), the scaling is $N[L]=185L-1115$ and
for a failure probability of 0.4 it becomes $N[L]\simeq16.6L-47.7$.
Now if we switch to our 3-qubit gate, then $L_c<2$ and our minimal
chain is now simply a 2-qubit cluster (locally equivalent to a Bell
state) yielding $N[L]=8L-44/3$. This is a vast improvement over
previous proposals.

For the two-qubit entangling gate, we essentially stand at the
same point as the photonic cluster state approaches. Optimizing
the resources boils down to finding the optimal strategies in
combining elements of cluster states. This is a very complex task, which
Gross \textsl{et al.} \cite{gro06} analyzed in great detail. 

For higher probabilities however, this critical length insuring average growth is no longer existent. All previously derived strategies become trivial within this probability range. Additional scalable approaches such as sequential adding are at hand and we shall go over the obvious ones. From previous works on generating cluster
states \cite{bar05a,dua05}, we know that the simplest way to grow short chains
with probabilistic gates is through a `divide and conquer' approach. It also turns
out to be much quicker than a sequential adding, as we allow for many gates to operate
in parallel. This technique links up chains of equal length on each round, and
discards the chains which failed to do so.

\begin{figure}
\begin{center}\includegraphics[scale=0.5]{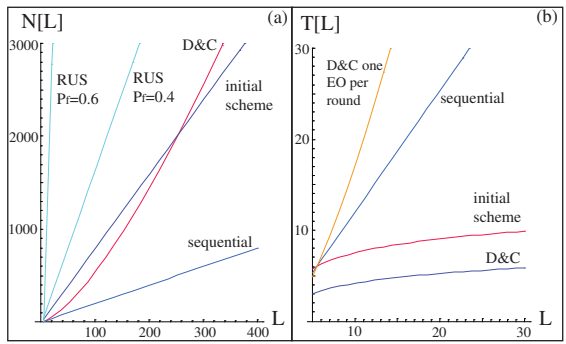}\end{center}
\caption{a) Comparison of entangling operation requirements for
chain production. We achieve much lower scalings in comparison with
those obtained through the repeat-until-success (RUS) scheme ($P_f$
being the failure probability). b) Time scaling for the different
strategies. The three-qubit gate is used in both plots.}
\label{fig5}
\end{figure}

In the context of higher success probabilities this approach can be extended to
growing large chains in the aim of saving time. The corresponding average number
of entangling operations becomes:
\begin{equation}
\label{32}
 N_{dc}[L]=\frac{(2/p)^{log_2(L-1)}-1}{2-p} \; .
\end{equation}
From the initial strategy we reach a value linear in $L$:
\begin{equation}
\label{33}
 N[L]=(2/p)\frac{L-1-2(1-p)/p}{1-2(1-p)/p}-1/p \; ,
\end{equation}
and a sequential adding yields:
\begin{equation}
\label{33a}
 N_{seq}[L]=(L-1)/(2p-1) \; .
\end{equation}
Obviously the latter represents a considerable saving, as can be verified in
Fig. (\ref{fig5}). Though the divide and conquer method doesn't scale linearly, up
till lengths of 250 qubits, it requires less entangling operations than the initial
scheme (which in fact is a full recycling approach). This is due to the fact that
the probabilities we are dealing with are significantly higher than in previous
proposals, which were undertaken in two steps, the building of minimal elements and
then their merging, in order to be scalable. If we look at the qubit resources however,
the less recycling we do, the more qubits we waste in the process. But as the success
probability of the gate increases, the recycling strategies all converge with the no-recycling
strategy (in terms of qubit resources), this being particularly noticeable for success
probabilities higher than 1/2.

We can also compare the time scaling of these various strategies, in units of time
$t$ corresponding to a single measurement. For the complete divide and conquer scheme
we simply have:
\begin{equation}
\label{34}
T_{dc}[L]=t\left(1+\log_2(L-1)\right) \; .
\end{equation}
and for the initial scheme:
\begin{equation}
\label{16b}
T[L]=(t/p)\left(1+\log_2\left(\frac{L-L_c}{L_0-L_c}\right)\right) \; .
\end{equation}
For the sequential adding, the cumulative time obeys $T_{L+1}=T_L+t/p$, and the
general form for $T$ becomes:
\begin{equation}
\label{17a}
T_{seq}[L]=t(L-1)/p \; .
\end{equation}
The first two approaches have a logarithmic dependence on the length $L$, however $T_{dc}$
is significantly lower as might have been expected (see Fig. (2)). Overall we see that
there is a clear advantage to divide the task up and to run parallel entangling operations.
The linear time scaling for the sequential method is due to the fact that operations cannot
be undertaken in parallel during the growth. If we didn't have access to simultaneous
entangling operations, the time scaling for the divide and conquer methods would be equivalent
to $N_{dc}[L]$ which is sub-exponential. One needs to keep in mind that by adopting a sequential
method, the whole procedure is simplified considerably and would be more accessible to physical
implementations.

\section{Discussion}

The cluster state comprises of active regions in which
it is being built or measured in the computation (both can be undertaken simultaneously)
and regions in which the qubits are simply waiting. Now this waiting can be minimized
in the building itself, through the appropriate protocols, and in the measurement
process. That is, the cluster can be built only a few layers in advance, so that the
qubits have less waiting to do, between the building and the actual measurement. In
any case, there will be some waiting. Therefore the lowest decoherence realization
would be preferred, but it may not be the easiest to manipulate. 
Thus we may envisage having
two different physical realizations constituting the cluster state.
For example, we could use single electron spins initially in building the cluster.
Once the links are made between one site and its nearest neighbors, the qubit
could be switched into a nuclear spin state which has a significantly longer
coherence time, via a swap operation or some other coherent write and read
actions. Most of the waiting would be done in the long-lived state, before
being swapped again for the readout \cite{kan98,loock06}. This follows the same
scenario as using a second physical system to mediate the interaction and make
the measurements, in distributed quantum computing. In the present proposal, we
use a continuous variable bus and homodyne measurements to generate the links. This
physical system shows itself to be very efficient in this application. Then, for
example, electron spins or superconducting charge qubits could be the matter
realization interacting with the bus and serving for the final readout. These systems
provide the useful static aspect required, they interact well with the mediating bus
and ensure good single qubit measurements. Finally a low decoherence realization
such as nuclear spin could be envisaged, mainly as a storage medium. The swapping
or write and read procedure should have a high fidelity for this storage to be
beneficial. On the whole, we see that optimization will depend directly on the
physical realization(s) we have chosen to work with. For systems with long dephasing
times we would give priority to sequential adding approaches, as we have some freedom
in the time scaling and thus we can make significant savings in resources. But for
realizations with short dephasing times, we would probably want to divide the task
up and run operations in parallel, in order to accelerate the fabrication of the
cluster state, at the expense of extra resources. 

\section{Conclusion}

In summary we have shown how the concept of the quantum bus can be
adapted to efficiently generating cluster states of matter qubits.
We can straightforwardly gain access to entangling probabilities higher than
1/2, removing the need to break up the building process into
inefficient and efficient parts. This opens up a new class of strategies, 
for which the resource consumption and the time scaling are consequently vastly improved. Clearly, within this class, detailed 
strategies can be envisaged and they will depend on the chosen physical
realization and the levels of decoherence present.

\section{Acknowledgements} We would like to thank R. van Meter and J. Eisert for helpful discussions. This
work was supported in part by MEXT in Japan and the EU project QAP.

\end{document}